\documentclass[11pt,english]{article}

\usepackage{graphicx}
\usepackage{dcolumn}
\usepackage{amssymb}
\usepackage{bm}
\usepackage{amsfonts}
\usepackage{amsmath}
\usepackage{bbm}
\usepackage[T1]{fontenc}
\usepackage[latin1]{inputenc}
\usepackage{babel}
\usepackage{color}

\begin{document}

\title{The Laser Gravitational Compass}

\author{ Ivan  S. Ferreira\\
University of Brasilia,  Institute of Physics, Brasilia, DF Brasilia, DF 70910-900, ivan@fis.unb.br,\vspace{2mm}\\
C. Frajuca\\
National Institute for Space Research, Sao Jose dos Campos, 12227-010, frajuca@gmail.com,\vspace{2mm}\\
 Nadja  S. Magalhaes\\
Physics Department,  Sao Paulo Federal University,  SP09913-030, Brazil, nadjasm@gmail.com,\vspace{2mm}\\
M. D. Maia\\
University of Brasilia,  Institute of Physics, Brasilia, DF70910-900,  maia@unb.br,\vspace{2mm}\\
Claudio M. G. Sousa\\
Federal University of Para, Santarem,  PA 68040-070, claudiogomes@ufpa.br.}

\maketitle

\begin{abstract}
 Using  the  observational properties of Einstein's   gravitational field  it is  shown  that a  minimum of four non-coplanar mass probes are   necessary   for  the  Michelson and Morley interferometer   to  detect gravitational waves within  the  context of  General Relativity.  With fewer probes,  some  alternative theories of gravitation can also explain the observations. The conversion of the existing   gravitational  wave detectors to  four probes  is also suggested.  
\end{abstract}

\section{The Observable Gravitational Wave}
Einstein's   prediction of gravitational waves (gw) was originally derived from  arbitrarily  small   perturbations of the  Minkowski metric  $g_{\mu\nu}= \eta_{\mu\nu} + h_{\mu\nu}$,  such  that   Einstein's  equations  reduce to  a linear wave  equation,  written in a special (de  Donder) coordinate  gauge,  as (Except when explicitly  stated,  Greek indices run from 0  to 3 and small case Latin indices run from 1 to 3).
\begin{equation}
\Box^2 \Psi_{\mu\nu} = 0, \;\;  \; \; \Psi_{\mu\nu}= h_{\mu\nu}-\frac{1}{2}h  \eta_{\mu\nu},\;\; \;\: h=\eta^{\mu\nu}h_{\mu\nu}. \label{eq:waves}
\end{equation}
The currently operating laser gw  observatories are inspired by  the Michelson \& Morley  (M\&M) interferometers\cite{Giazotto,Pitkin,Eardley},  where the data acquired  by the Fabri-Per\'ot interferometer (an etalon)   is  used to generate a    numerical simulation,  thus  producing  a  template from which the most probable source  is estimated\cite{Abbot2016}.
 The purpose of this note is  to show that in order to detect  gw described by  General Relativity (GR) with  an  M\&M  interferometer  requires  a minimum of  four non-coplanar mass probes. 

The observables of  Einstein's   gravitational  field  are   given  by  the  eigenvalues of the  Riemann curvature\cite{Zakharov,Pirani1956,Pirani1957},   defined by  $R(U,V)W = [\nabla_U, \nabla_V]W - \nabla_{[U,V]}W$,  whose components in  any  given basis  $ \{e_\mu \} $   are  $R(e_\mu,e_\nu)e_\rho = R_{\mu\nu\rho\sigma}e^\sigma$. Then we  find that there are at most six independent eigenvectors $X_{\mu\nu}$  and   six  eigenvalues $\lambda$,  solutions of the  eigenvalue  equation  $ R_{\mu\nu\rho\sigma}X^{\rho\sigma}=\lambda X_{\mu\nu}$, including the  zero  eigenvalue, corresponding to the  absence of gravitation\cite{Pirani1962,Sachs1962a,Pirani1967}.  Thus, using  the language of field theory,  Einstein's gravitation is  said  to have  five non-trivial  observables or degrees  of  freedom  $(dof)$. The spin-statistics theorem  relates the  $dof$ to  the helicity or the orbital spin of  the field as  $s=(dof-1)/2$, so that  Einstein's  gravitation is also  said  to be  a  spin-2  field.   

Alternative  theories of gravitation may have distinct definitions of  observables (not necessarily related to  curvature) and  their gravitational waves, if  they exist,  may require  different methods of observations.  Well known   examples include:  the  spin-1 gauge theories of  gravitation (there are  several of them),  characterized by  dof=3; topological   gravitation in three dimensions; projective   theories  of  gravitation; F(R) theories; among   many others.  Therefore, in order to  understand  the  observation of  a  gw  it is  essential  to  specify  the observables of  the theory on which the  experiment is based. 

The most general massless spin-2 field  $h_{\mu\nu}$  was  defined   in the Minkowski space-time  by  Fierz  and Pauli\cite{FierzPauli},  as a   trace-free field   $h=h_\mu{}^\mu=0$,  satisfying the field  equations  $\Box^2 h_{\mu\nu}=0$.  This   is a  linear field  not to be  confused with Einstein's  gravitation.  Since in the case of the present  observations  of  gw, the  supporting theory is Einstein's  gravitation,  then  the  observational signature  to be sought  is that of a  $dof=5$ or of a spin-2  field, characterized by the   observable  curvature. 

The use of the Fierz-Pauli field  $h_{\mu\nu}$ as the perturbation of the Minkowski metric  makes it possible to  free Eq. (\ref{eq:waves}) of coordinate gauges   so that its    solutions  can be  written as  as a superposition of plane  polarized gravitational waves, characterized by  the  Traceless-Transverse-Plane-Polarized (TTPP) gauge  conditions\cite{Giazotto}:
\begin{equation}
h=0,\;\;  h_{i0}=0, \;\; \Box^2 h_{\mu\nu}=0,\;\;  h_{\mu\nu;\rho}=0. \label{eq:TTPP}
\end{equation}
Then,  these  conditions are   used  to  simulate  a  template, from  which the  source source  of the  gw observations  is  estimated.
 
\section{The Equivalence Principle and  the  M\&M gw  Detector}	
 The  use   of  an  M\&M detectors for  gw is  based on  the  principle  of  equivalence of GR:   Given  2  masses A  and  B with attached  mirrors,  under the exclusive action of a  known gravitational  field,   they  propagate (or ``free fall'') along  time-like geodesics,  with unit  tangent vectors  $T_A$ and  $T_B$  respectively, satisfying  geodesic  equations  $\nabla_{T_{A}} T_A=0$  and $\nabla_{T_{B}} T_B=0$ . Eventually,  probe  A sends a light signal with velocity $P$ to particle B   along  the  light geodesic with equation $\nabla_P P=0$. After a  while, probe  A  receives back  the reflected signal,  so that  these  geodesics  describe a closed parallelogram,  with the  closing condition   $\nabla_T P=\nabla_P T$.

The  curvature  tensor calculated  in that  parallelogram is $ R(T,P)T = [\nabla_T ,\nabla_P ]T =  \nabla_T (\nabla_T P)  = \nabla_T a$, where $a = \nabla_T P$  is the  acceleration of the signal along the fall.  Defining a basis   by  $\{e_0 =T , e_i= P \} $, where $e_i$  denotes  any space-like  direction,   we  obtain  the  geodesic  deviation  equation in   components $\frac{d a_i}{cdt}=R_{0i0i}$,  where $t$ denotes the time parameter of the time-like  geodesic.   As we can see,  the  motion  of the probes  generates a  2-dimensional world-sheet, whose  curvature $R_{0i0i}$, is translated directly  into the  the variation  of the fall acceleration.    

In the  currently operating and  some  future  planned  M\&M  gw detectors,   three   mass-probes  are  used,  defining a  space-like plane in space-time,   whose   motion under the action of  a gw generates  a 3-dimensional  world-volume,  whose curvature is  measured by a  $2 \times 2$   acceleration array  $a_{ij}$, obeying the geodesic deviation equation
\begin{equation}
\frac{d a_{ij}}{cdt}= \!\!R_{0i0j}, \; i,j =1,2.   \label{eq:GDE2}
\end{equation}
Therefore,   such detectors  are   capable  to measure   only 3 curvature  components  $R_{0101},R_{0102},R_{0202}$, so that at most  three  degrees of  freedom  of  Einstein's gravitational field  can be obtained.  Since the  observed gw are  very weak,  it has been  assumed that the  missing   degrees of freedom in one detector may be  complemented   by  the data collected  on another  detector located  somewhere  else. Such understanding  is  supported by the  numerical simulation, from which  an estimate of   the wave source capable of reproducing the   same  observation  in two separate  detectors  is  obtained.  Although   it  is possible  to  parallelly  transport  the curvature  tensor  from  one detector  to  another,  the left  hand  side content of   Eq. (\ref{eq:GDE2})  represents a locally measured quantity which  cannot  be transported  to another detector,  under the penalty of  breaching the principle of  equivalence.
 
The definitive  solution for the   missing   dof problem  can be obtained  by a direct   measure of   the curvature tensor using the  geodesic  deviation  equations Eq. (\ref{eq:GDE2}),  extended to   $i,j =1,2,3$.  One such  detector, the  ``gravitational compass'' was  conceived by   P. Szekeres. It  consists of four  non-coplanar   mass probes connected by  six  dynamometers,  to measure  the  six curvature eigenvectors. By comparing  three of these  eigenvectors with  the three principal  directions of  curvature,  the  compass can  points  directly to the  source\cite{Szekeres}. 
 
Szekeres'  gravitational   compass was  improved   by F. Pirani\cite{Pirani1967}, where instead  the  dynamometers, the four mass-probes had attached  mirrors and  light sources. Then,  applying  Eq. (\ref{eq:GDE2})  again for  $i,j =1,2,3$,   all components of the  curvature  can be read locally,  defining the 4-dimensional gravitational field  at the center of mass of  the  four  probes.  The interesting fact is  that  four  non-coplanar masses define  a  virtual  spheroidal surface in  space-time, then the  natural oscillation modes of the  spheroid  give a  direct measure  of the  curvature eigenvalues.     

 The  Szekeres-Pirani gravitational compass can  be  implemented in the presently operating  gw  observatories,  with the  addition of  a  fourth  mass probe not belonging to  the  plane of the  existing  2-dimensional M\&M detectors,  located either in a   tower or well, not necessarily having the  same  length  as  the  existing arms,  thus defining  a spheroidal  gw  detector.  
\vspace{2mm}\\
Summarizing,   the  presently operating gw  detectors are hindered by the fact that  they  cannot  detect directly the all degrees of  freedom of  Einstein's   gravitational  field at  each point,  thus leaving  open the   possibility of  alternative explanations  for the  gw.  For the specific case of gravitation in the sense of  Einstein's theory,  we  suggest   the  addition of  a fourth  mass to  the existing  detectors,  so  that all degrees of  freedom of  gravitation are directly  measured from  the  five fundamental  oscillation modes of  a  spheroid.

 \textbf{Acknowledgments}: \small NSM and CF acknowledge FAPESP for its support to their research through the thematic project 2013/26258-4.

\end{document}